\documentclass[%
 reprint,
 amsmath,amssymb,
 aps,
]{revtex4-1}

\usepackage{graphicx}
\usepackage{dcolumn}
\usepackage{bm}
\usepackage{color}
\usepackage{soul}

\begin{document}


\title{Tricritical behavior of soft nematic elastomers}

\author{Danilo B. Liarte}
\email[]{dl778@cornell.edu}
\affiliation{Laboratory of Atomic and Solid State Physics, Clark Hall, Cornell University, Ithaca, New York 14853-2501, USA}
\affiliation{Intituto de F\'isica, Universidade de S\~ao Paulo \\ Caixa Postal 66318, CEP 05314-970 S\~ao Paulo, SP, Brazil}

\date{\today}

\begin{abstract}
We propose a lattice statistical model to investigate the phase diagrams and the soft responses of nematic liquid-crystal elastomers. Using suitably scaled infinite-range interactions, we obtain exact self-consistent equations for the tensor components of the nematic order parameter in terms of temperature, the distortion and stress tensors, and the initial nematic order. These equations are amenable to simple numerical calculations, which are used to characterize the low-temperature soft regime. We find a peculiar phase diagram, in terms of temperature and the diagonal component of the distortion tensor along the stretching direction, with first- and second-order transitions to the soft phase, and the prediction of tricritical points. This behavior is not qualitatively changed if we use different values of the initial nematic order parameter.
\end{abstract}

\maketitle

\section{Introduction}

Since de Gennes' pioneering work in 1969 \cite{gennes69}, liquid crystal elastomers have continually interested theoreticians, in part due to their intriguing mechanical and orientational behavior. Although much knowledge has been accumulated, specially from the application of continuum theories \cite{stenull04, ennis06, ye07, zhu11} and numerical simulations \cite{pasini05, skacej11, whitmer13}, there are just a few insights from microscopic statistical models \cite{selinger04, xing08, liarte11}. The neoclassical theory of elasticity is a step forward, as it provides a statistical basis for the elastic properties of nematic elastomers (NEs), and has been successfully applied to a number of phenomena \cite{bladon94, warner96, warner03}. This approach, however, is based on a global average tensor for the nematic order parameter, and hence cannot be strictly regarded as a microscopic calculation. In a previous paper, we developed a mean-field approach \cite{liarte11} based on a statistical model used by Selinger and Ratna \cite{selinger04}, which combines the neo-classical theory and a lattice version of the Maier-Saupe theory of the isotropic-nematic transition \cite{lebwohl72}. The present work is an outgrowth of our previous treatment. We now look at the soft response and the soft transitions of NEs.

\begin{figure}[!ht]
\vspace{0.5cm}
 \begin{center}
 \includegraphics[width=0.9\linewidth]{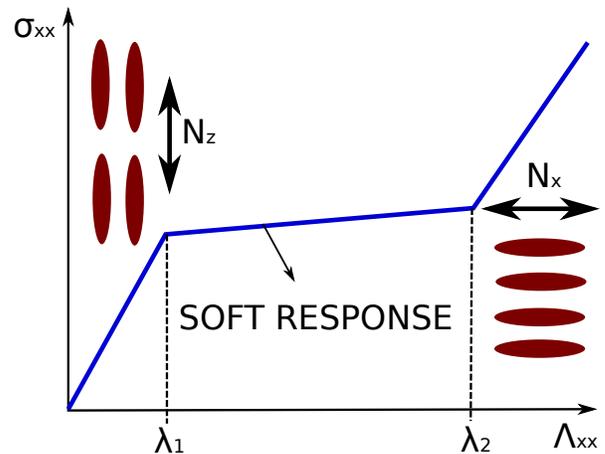}
 \end{center}
 \caption{(Color online) Schematic plot of the $xx$-components of the distortion tensor versus the engineering stress tensor.}
 \label{softresponse}
\end{figure}

It is known that NEs may exhibit soft response. One of the typical features of soft behavior is the response of stress to distortion of the sample, as we sketch in Fig. \ref{softresponse}. Consider a sample of a nematic elastomer that has been initially cross-linked in the nematic state, with orientational order along the anisotropic $z$-axis, for example. In many experiments, the nominal stress is measured as a function of an applied uniaxial deformation along a direction perpendicular to this anisotropy axis (the $x$ axis, for example). Let us call $\lambda$ the $xx$-component of the distortion tensor, and $\sigma$ the $xx$-component of the nominal stress. Soft response is observed at the almost level region between $\lambda_{1}$ and $\lambda_{2}$ in Fig. \ref{softresponse} \footnote{The usual distinction between ``soft'' and ``semi-soft'' responses associates the later with the existence of Hookean behavior at the beginning of the stress-strain curve.}. If it is plotted versus strain, the stress presents a rather flat plateau in the soft region. This plateau is related to a vanishing of the elastic modulus that measures the energy associated with shear in the $xz$-plane \cite{ye07}. It has been interpreted in terms of rotations of the distribution of shapes of the anisotropic chains without distortion \cite{warner02}. This plateau has also been associated with a genuine phase transition \cite{ye09}, with the spontaneous emergence of a non-vanishing shear component of the strain tensor along the $xz$-plane. In fact, the $xz$-component of the strain increases from zero at $\lambda_{1}$, reaches a maximum value, and then decreases to zero again at $\lambda_{2}$. The nematic phase is uniaxial for $\lambda<\lambda_{1}$ (along the $z$ direction) and for $\lambda>\lambda_{2}$ (along the $x$-direction).

\begin{figure}[!ht]
\vspace{0.5cm}
 \begin{center}
 \includegraphics[width=\linewidth]{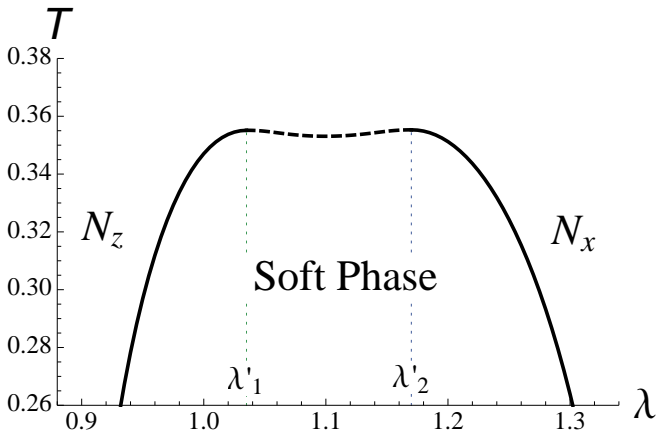}
 \end{center}
 \caption{Phase diagram in terms of the $xx$-component of the distortion tensor $\lambda$ and temperature $T$. $N_x$ and $N_z$ denote nematic uniaxial phases along the $x$ and $z$-axes. The shear component of the distortion tensor along the $xz$-plane is zero in $N_x$ and $N_z$, and nonzero in the soft phase. The two tricritical points at $\lambda_1^\prime$ and $\lambda_2^\prime$ separate first (dashed) and second order transition lines.}
 \label{phasediagram1}
\end{figure}

In this paper we investigate the static properties of a mean-field statistical model to describe the soft transition in nematic elastomers. Our most important result is the phase diagram of Fig. \ref{phasediagram1}, in terms of the $xx$-component of the distortion tensor $\lambda$ and temperature $T$ . We characterize a uniaxial and a soft phase \footnote{There is no transition between Nz and Nx. They belong to the same standard uniaxial phase. The optical axis continuously rotates from the z- to the x-direction upon increasing $\lambda$. Also, there is no nematic-isotropic transition, since the imposed strain acts as an aligning mechanical field.}. The standard uniaxial phase (with the nematic anisotropy changing from $z$ to $x$ directions) is associated with no shear along the $xz$-plane. The soft phase is still uniaxial nematic, but displays a finite shear strain component along the $xz$-plane. The solid and dashed lines represent second and first-order transitions to the soft phase respectively. The two continuous transition lines meet the coexistence line at two tricritical points, at $\lambda_{1}^{\prime}$ and $\lambda_{2}^{\prime}$. This general qualitative behavior does not depend on the initial nematic order. In the next section, we use some ideas of our previous approach \cite{liarte11} to construct a more complete model for the nematic elastomers. In Section III, we describe some of the steps to obtain the analytical solutions of this problem in the mean-field approximation. In section IV, we study the numerical solutions of the self-consistent equations and discuss the most important results. A summary of these results is presented in the last Section.

\section{Model}

In this approach, the principal axes of uniaxial mesogens are associated with a set of vectors ($\{\bm{n}_i\}$, $i=1, \cdots, N$) defined in the unit sphere ($|\bm{n}_i|=1$). We consider a soft quadrupolar form \cite{maier58,lebwohl72} for the energy of interaction between mesogens,
\begin{eqnarray} 
E_{soft} = -A \sum_{\langle i,j \rangle} \sum_{\mu,\nu \in \{x,y,z\}} S_i^{\mu \nu} S_j^{\mu \nu},
\label{esoft}
\end{eqnarray}
where the first sum is restricted to pairs of nearest neighbors of an arbitrary crystalline lattice \footnote{With appropriate choices for $A$ and the zero-point energy, this model can be mapped into the Lebwohl-Lasher model \cite{lebwohl72}. Later on we will take the mean-field approximation by considering infinite-range interactions between mesogens \cite{kac68,stanley71,salinas93}, so that the initial lattice topology becomes irrelevant.}. The uniaxial quadrupole moments may be written in terms of the local vector components ($n_i^\mu$, $\mu\in\{x,y,z\}$) as
\begin{eqnarray}
S_i^{\mu \nu} = \frac{1}{2} \left(3 \, n_i^\mu n_i^\nu - \delta^{\mu \nu} \right),
\label{smunu}
\end{eqnarray}
where $\delta^{\mu \nu}$ is the Kronecker delta.

Deformations are described by the distortion tensor $\Lambda$, with components $\Lambda_{\mu \nu} = \partial R_\mu/\partial R_{0, \nu}$, where $R_\mu$ and $R_{0, \mu}$ are the spatial coordinates of a point in the reference and target spaces, respectively \cite{landau86,chaikin95}. In this paper we consider a coarse-grained homogeneous distortion tensor. The homogeneity assumption is made for mathematical simplicity. It leaves out the possibility of emergent micro-structured behavior, as discussed in the Conclusions section. Nonetheless, the model lends itself to in-depth analytical calculations. We also assume that the elastic part of the system's free energy arises exclusively from entropic effects, so that rubber elastic effects can be approximated by a degeneracy factor in the partition function,
\begin{eqnarray}
Z = \sum_{\{\bm{n}_i\}} \Omega \left(\{\bm{n}_i\},\Lambda \right) e^{-\beta E_{\text{soft}}}.
\end{eqnarray}
Accordingly, this degeneracy factor \cite{selinger04,liarte11} may be associated with the ``trace formula'' derived in Warner-Terentjev theory of elasticity \cite{warner03}, 
\begin{eqnarray}
\Omega = \exp \left[-\frac{n_s}{2} \sum_{i=1}^N \text{Tr} \left(l_{0,i} \cdot \Lambda^T \cdot l_i^{-1} \cdot \Lambda \right) \right],
\end{eqnarray}
where $n_s$ is the number of strands per unit volume. $l_i$ is an effective shape tensor, and $l_{0,i}$ is an effective shape tensor at the time of cross-linking, and may be written in terms of the quadrupole moments as
\begin{eqnarray}
l_{i, \alpha \beta}^{-1} = a \left(\delta^{\alpha \beta} - b S_i^{\alpha \beta}\right),
\end{eqnarray}
\begin{eqnarray}
l_{0 i, \alpha \beta}^{-1} = a \left(\delta^{\alpha \beta} - b S_{0,i}^{\alpha \beta}\right),
\end{eqnarray}
where $a$ and $b$ are positive constants \footnote{Notice that $b=0$ for isotropic rubber. Also, $a=(2l_{\perp}^{-1} + l_{\parallel}^{-1})/3$, and $b=2(l_{\parallel}^{-1}-l_{\perp}^{-1})/3$, in terms of the parallel ($l_\parallel$) and perpendicular ($l_{\perp}$) effective displacements of the polymer chain \cite{liarte11,warner03}. Information about the type of nematic polymer (main chain and side chain) is condensed in these parameters. In general, the ratio $l_{\parallel} / l_{\perp}$ is higher for main chain nematic polymers, leading to more extreme mechanical effects (see section 3.2.1 of \cite{warner03})}, and we have introduced the components of the local nematic order tensor $S_{0,i}^{\alpha \beta}$ at the time of cross-linking.

We consider $\{S_{0,i}\}$ as a set of quenched i.i.d. random variables, satisfying the discrete probability distribution,
\begin{eqnarray}
S_{0,i} =
\left\{ \begin{array}{lll}
S_x, &\text{with probability}& 1/3-p/2, \\
S_y, &\text{with probability}& 1/3-p/2, \\
S_z, &\text{with probability}& 1/3+p,
\end{array} \right.
\label{discreteS0}
\end{eqnarray}
where
\begin{eqnarray}
S_x&=&\text{Diag}\left[2,-1,-1\right], \\
S_y&=&\text{Diag}\left[-1,2,-1\right], \\
S_z&=&\text{Diag}\left[-1,-1,2\right].
\end{eqnarray}
Notice that there is a preferred orientational ordering along the $z$-axis for $p\in \left[0,2/3\right]$. Also, according to the law of large numbers,
\begin{eqnarray}
\lim_{N \rightarrow \infty} \frac{1}{N} \sum_{i=1}^N S_{0,i}^{\alpha \beta} = \langle S_{0,i}^{\alpha \beta} \rangle = \frac{S_0}{2} \left(\begin{array}{ccc}
-1&0&0 \\
0&-1&0 \\
0&0&2
\end{array}\right),
\end{eqnarray}
where $S_0$ is the scalar nematic order parameter at the time of cross-linking, and the last equality follows from a standard parametrization for uniaxial nematic systems. Now we can relate $p$ and $S_0$ through
\begin{eqnarray}
p=\frac{2}{3} S_0,
\end{eqnarray}
which turns out to be a convenient way of incorporating the reference space anisotropy into the theory. We call the attention to the fact that our choice for the simple discrete distribution of Eq. \eqref{discreteS0} can be easily replaced by more realistic model-specific distributions. After some algebra, the local shape tensor at the time of cross-linking may be written as
\begin{eqnarray}
l_{0,i}^{\alpha \beta} = a^{-1} \left(\delta^{\alpha \beta}+bL_i^{\alpha \beta}\right),
\label{l0rf}
\end{eqnarray}
where we have defined random-field-like variables $L_i$, which satisfy the probability distribution,
\begin{eqnarray}
L_i=\left\{
\begin{array}{lll}
H_x, &\text{with}& \mathcal{P}_x=(1-S_0)/3, \\
H_y, &\text{with }& \mathcal{P}_y=(1-S_0)/3, \\
H_z, &\text{with}& \mathcal{P}_z=(1+2S_0)/3,
\end{array} \right.
\end{eqnarray}
with
\begin{eqnarray}
H_x&=&\text{Diag}\left[(1-b)^{-1},-(2+b)^{-1},-(2+b)^{-1}\right], \\
H_y&=&\text{Diag}\left[-(2+b)^{-1},(1-b)^{-1},-(2+b)^{-1}\right], \\
H_z&=&\text{Diag}\left[-(2+b)^{-1},-(2+b)^{-1},(1-b)^{-1}\right].
\end{eqnarray}
Now we can plug \eqref{l0rf} back in the trace formula to write
\begin{eqnarray}
\ln \Omega&=& -\frac{n_s}{2} \sum_{i=1}^N \left[\text{tr} \left(\Lambda^T \cdot \Lambda\right) + b\, \text{tr} \left(L_i \cdot \Lambda^T \cdot \Lambda \right) \right. \nonumber \\ && \quad \left. - b\, \text{tr} \left(\Lambda^T \cdot S_i \cdot \Lambda \right) - b^2 \text{tr} \left(L_i \cdot \Lambda^T \cdot S_i \cdot \Lambda \right) \right].
\end{eqnarray}
We can also associate a set of matrices $S_i$ with the tensorial components $S_i^{\mu \nu}$, so that
\begin{eqnarray}
E_{\text{soft}} = -A \sum_{\langle i, j \rangle} \text{tr} \left( S_i \cdot S_j \right).
\end{eqnarray}
The partition function is then given by
\begin{eqnarray}
Z &=& \sum_{\{S_i\}} \exp \left\{
\beta A \sum_{\langle i, j \rangle} \text{tr} \left( S_i \cdot S_j \right) - \frac{n_s}{2} \sum_{i=1}^N 
\right. \nonumber \\ && \quad \left.
\text{tr}\left[\Lambda^T \cdot \Lambda
+ b\,  L_i \cdot \Lambda^T \cdot \Lambda - b\, \Lambda^T \cdot S_i \cdot \Lambda \right. \right. \nonumber \\ && \quad \left. \left. - b^2 L_i \cdot \Lambda^T \cdot S_i \cdot \Lambda \right]
\right\},
\end{eqnarray}
and the free energy,
\begin{eqnarray}
f&=&f_{\text{rub}} - kT \lim_{N\rightarrow \infty} \frac{1}{N} \ln \left\{
 \exp \left[-\frac{n_s b}{2} \sum_i \text{tr} \left(L_i 
 \right. \right. \right. \nonumber \\ && \quad \left. \left. \left. \cdot
 \Lambda^T \cdot \Lambda \right)\right]
 \sum_{\{S_i\}} \exp \left[
\beta A \sum_{\langle i, j \rangle} \text{tr} \left( S_i \cdot S_j \right)+ \frac{n_s b}{2} 
\right. \right. \nonumber \\ && \quad \left. \left. \times
\sum_i \text{tr} \left(\Lambda^T \cdot S_i \cdot \Lambda \right)
+ \frac{n_s b^2}{2} \sum_i \text{tr} \left(L_i 
\right. \right. \right. \nonumber \\ && \quad \left. \left. \left.
\cdot \Lambda^T \cdot S_i \cdot \Lambda \right)
\right] \right\},
\end{eqnarray}
where
\begin{eqnarray}
f_{\text{rub}} = \frac{n_s k T}{2} \text{tr} \left(\Lambda^T \cdot \Lambda\right),
\end{eqnarray}
is the free energy of the isotropic rubber.

\section{Mean-field calculations}

Exact analytical results at the mean-field level can be obtained by considering a simplified model with adequately scaled fully-connected interactions \cite{kac68,stanley71,salinas93}. In the present approach, we replace soft quadrupole interactions by what we call Maier-Saupe model \cite{henriques97, carmo10, liarte11, liarte12},
\begin{eqnarray}
E_{MS} = -\frac{A}{2N} \sum_{i,j=1}^N S_i^{\mu \nu} S_j^{\mu \nu},
\label{ems-mf}
\end{eqnarray}
where the scale with $N$ ensures the existence of a well-defined thermodynamic limit. In contrast with our previous approach \cite{liarte11}, we will not use the Zwanzig approximation in this work \cite{zwanzig63}, since we have not found a globally stable soft solution for the discrete model. The full-continuum model turns out to be computationally costly though. In order to linearize the quadratic form in Eq. \eqref{ems-mf}, we use a set of Gaussian identities,
\begin{eqnarray}
&& \exp\left[\frac{\beta A}{2N} \sum_{\mu, \nu} \left(\sum_i S_i^{\mu \nu}\right)^2\right] = 
\nonumber \\ && \quad
\int d\mu_Q e^{-N\beta A \text{tr} Q^2 / 2} \exp \left[\beta A \sum_i \text{tr} \left(Q \cdot S_i \right)\right],
\end{eqnarray}
where $d\mu_Q = \prod_{\mu,\nu} \sqrt{\beta A N/2 \pi} \, dQ_{\mu \nu}$. Thus,
\begin{eqnarray}
f&=&f_{\text{rub}} - kT \lim_{N\rightarrow \infty} \frac{1}{N} \ln \left\{
 \exp \left[-\frac{n_s b}{2} \sum_i \text{tr} \left(L_i 
 \right. \right. \right. \nonumber \\ && \quad \left. \left. \left.
 \cdot \Lambda^T \cdot \Lambda \right)\right] \int d\mu_Q e^{-N\beta A \text{tr} Q^2 / 2}  \sum_{\{S_i\}} \left[
\exp \left(\beta A
\right. \right. \right. \nonumber \\ && \quad \left. \left. \left.
\times \sum_i \text{tr} \left(Q \cdot S_i \right)
+ \frac{n_s b}{2} \sum_i \text{tr} \left(\Lambda^T \cdot S_i \cdot \Lambda \right)
\right. \right. \right. \nonumber \\ && \quad \left. \left. \left.
+ \frac{n_s b^2}{2} \sum_i \text{tr} \left(L_i \cdot \Lambda^T \cdot S_i \cdot \Lambda \right)
\right) \right] \right\}.
\end{eqnarray}
Now we are able to write the sum over states as
\begin{eqnarray}
&& \prod_i \left\{\int_{\mathcal{S}^2} \exp \left[\beta A \, \text{tr}\left(Q \cdot S\right) + \frac{n_s b}{2} \text{tr} \left(\Lambda^T \cdot S \cdot \Lambda \right) 
\right. \right. \nonumber \\ && \left. \left.
+\frac{n_s b^2}{2} \text{tr} \left(L_i \cdot \Lambda^T \cdot S \cdot \Lambda \right)
\right]
\right\},
\end{eqnarray}
where $\int_{\mathcal{S}^2}=\int_0^\pi \sin \phi d\phi \int_0^{2\pi} d\theta$ denotes an integral over the surface of a unit sphere. After some algebra, the free energy reads
\begin{eqnarray}
f&=&f_{\text{rub}} +\frac{n_s k T b}{2} \sum_{\mu \in \{x,y,z\}} \mathcal{P}_\mu \text{tr} \left(H_\mu \cdot \Lambda^T \cdot \Lambda \right)
\nonumber \\ &&
-kT \lim_{N\rightarrow \infty} \frac{1}{N} \ln \left\{
\int d\mu_Q \exp \left[
-\frac{N \beta A}{2} \text{tr} Q^2 
\right. \right. \nonumber \\ && \left. \left.
+ N \sum_\mu \mathcal{P}_\mu \ln \left(
\int_{\mathcal{S}^2} \exp \left(\text{tr} \left(
\left(
\beta A Q + \frac{n_s b}{2} \Lambda \cdot \Lambda^T
\right. \right. \right. \right. \right. \right. \nonumber \\ && \left. \left. \left. \left. \left. \left.
+\frac{n_s b^2}{2} \Lambda \cdot H_\mu \cdot \Lambda^T \right) \cdot S
\right)
\right)
\right)
\right]
\right\},
\label{fen-lln}
\end{eqnarray}
where we have applied the law of large numbers to simplify the random-field interaction term. In the limit of large $N$, the integral over the $Q$-variables in Eq. \eqref{fen-lln} may be evaluated by Laplace's method, so that \footnote{Whenever possible we use Einstein summation rule for repeated indices in order to simplify notation.},
\begin{eqnarray}
f&=&f_{\text{rub}} +\frac{n_s k T b}{2} \mathcal{P}_\mu \text{tr} \left(H_\mu \cdot \Lambda^T \cdot \Lambda \right)+\frac{A}{2} \text{tr} Q^2 
\nonumber \\ &&
-kT  \mathcal{P}_\mu \ln \left\{
\int_{\mathcal{S}^2} \exp \left[\text{tr} \left(
\left(
\beta A Q + \frac{n_s b}{2} \Lambda \cdot \Lambda^T
\right. \right. \right. \right. \nonumber \\ && \left. \left. \left. \left.
+\frac{n_s b^2}{2} \Lambda \cdot H_\mu \cdot \Lambda^T \right) \cdot S
\right)
\right]
\right\},
\label{fen-a}
\end{eqnarray}
where the order parameter components satisfy the set of self-consistent equations,
\begin{eqnarray}
Q_{\alpha \beta} = \sum_\mu \mathcal{P}_\mu \displaystyle\frac{\displaystyle\int_{\mathcal{S}^2} S_{\alpha \beta} e^{G_\mu(Q,\Lambda, H_\mu)}}{\displaystyle\int_{\mathcal{S}^2}  e^{G_\mu(Q,\Lambda, H_\mu)}},
\end{eqnarray}
with
\begin{eqnarray}
G_\mu &=& \text{tr} \left[\left(\beta A Q + \frac{n_s b}{2} \Lambda \cdot \Lambda^T + \frac{n_s b^2}{2} 
\right. \right. \nonumber \\ && \quad \left. \left.
\times \Lambda \cdot H_\mu \cdot \Lambda^T \right) \cdot S\right].
\end{eqnarray}

\section{Results}

Henceforth we consider dimensionless variables $f \leftrightarrow f/A$, $kT/A \leftrightarrow T$ to simplify the equations, so that
\begin{eqnarray}
f &=& \frac{n_s T}{2} \text{tr} \left[\left(\delta + b \mathcal{P}_\mu H_\mu \right) \cdot \Lambda^T \cdot \Lambda \right] + \frac{1}{2} \text{tr} Q^2 
\nonumber \\ && \quad
- T p_\mu \ln \left(\int_{\mathcal{S}_2} e^{G_\mu}\right),
\end{eqnarray}
and,
\begin{eqnarray}
G_\mu = \text{tr} \left\{\left[\frac{1}{T} Q + \frac{n_s b}{2} \Lambda \cdot \left(\delta + b H_\mu \right) \cdot \Lambda^T\right] \cdot S\right\}.
\end{eqnarray}

We consider the following parametric form for the global distortion tensor in cartesian coordinates,
\begin{eqnarray}
\Lambda = \left(\begin{array}{ccc}
\lambda & 0 & \kappa \\
0 & 1 & 0 \\
0 &0 & 1/\lambda
\end{array}
\right),
\label{dist-tensor}
\end{eqnarray}
which describes incompressible deformations ($\text{det} \Lambda = 1$) with a shear component in the $xz$-plane. In our numerical analysis, we will keep $\lambda$ fixed, and study the dependence of the free energy on $\kappa$. In the stationary point, $\partial f / \partial \kappa = 0$ implies
\begin{eqnarray}
&& \frac{\partial}{\partial \kappa} \left\{\frac{n_s T}{2} \text{tr} \left[\left(\delta + b \mathcal{P}_\mu H_\mu \right)\cdot \Lambda^T \cdot \Lambda\right] \right\} 
\nonumber \\ && \quad
-T \sum_\mu \mathcal{P}_\mu \displaystyle\frac{\displaystyle\int_{\mathcal{S}^2} \left(\frac{\partial G_\mu}{\partial \kappa}\right)e^{G_\mu}}{\displaystyle\int_{\mathcal{S}^2} e^{G_\mu}} =0.
\end{eqnarray}
The first Piola-Kirchhoff, or engineering, stress tensor \cite{marsden68,warner03} is defined by
\begin{eqnarray}
\sigma_{\mu \nu} = \frac{\partial f}{\partial \Lambda_{\mu \nu}}.
\end{eqnarray}
The equation for $\sigma_{xx}=\sigma$ is then given by
\begin{eqnarray}
\sigma &=& \frac{\partial}{\partial \lambda} \left\{\frac{n_s T}{2} \text{tr} \left[\left(\delta + b \mathcal{P}_\mu H_\mu \right)\cdot \Lambda^T \cdot \Lambda\right] \right\} 
\nonumber \\ && \quad
-T \sum_\mu \mathcal{P}_\mu \displaystyle\frac{\displaystyle\int_{\mathcal{S}^2} \left(\frac{\partial G_\mu}{\partial \lambda}\right)e^{G_\mu}}{\displaystyle\int_{\mathcal{S}^2} e^{G_\mu}}.
\end{eqnarray}
At last, the order parameter tensor $Q$ may be represented by the traceless symmetric matrix,
\begin{eqnarray}
Q = \left(\begin{array}{ccc}
Q_{xx} & 0 & Q_{xz} \\
0 & -\left(Q_{xx}+Q_{zz}\right) & 0\\
Q_{xz} & 0 & Q_{zz}
\end{array}
\right),
\end{eqnarray}
where $Q_{yx}=Q_{yz}=0$ by symmetry arguments.

\begin{figure*}[!ht]
\vspace{0.5cm}
\begin{minipage}[b]{0.48\linewidth}
\includegraphics[width=\linewidth]{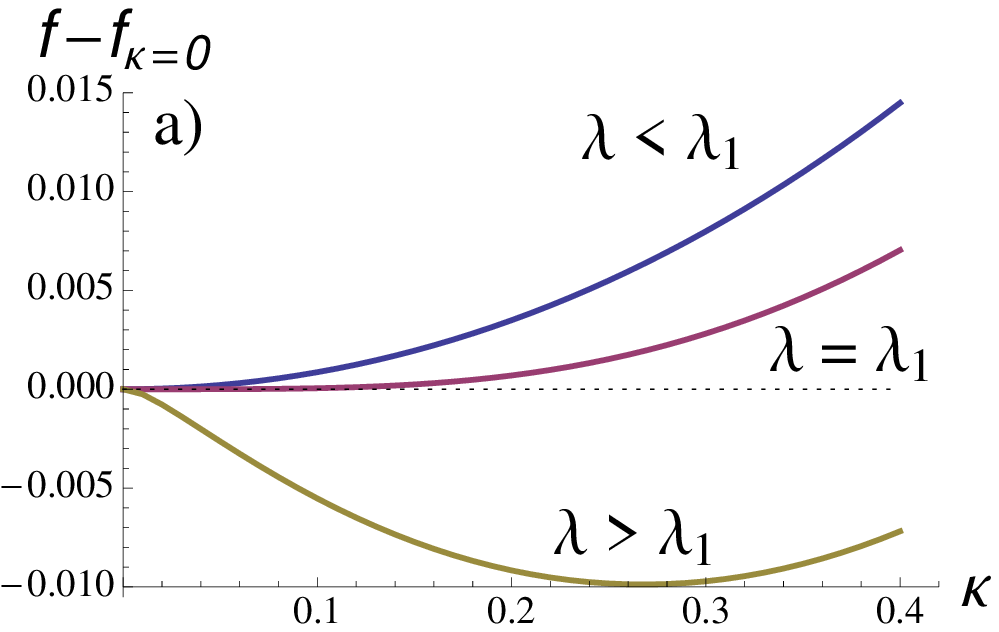}
\end{minipage}
\hspace{0.5cm}
\begin{minipage}[b]{0.48\linewidth}
\includegraphics[width=\linewidth]{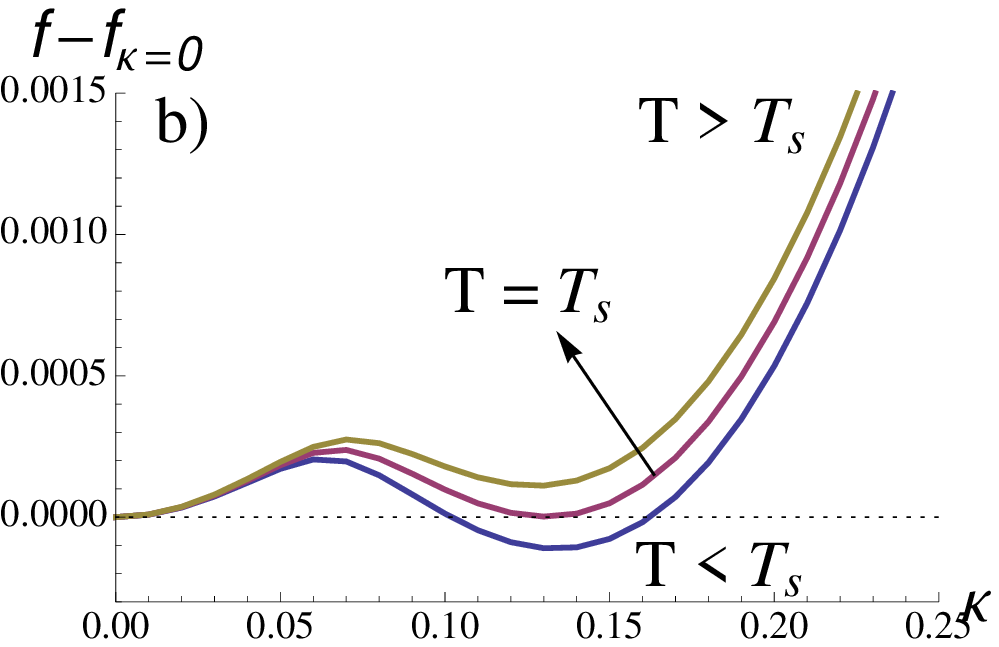}
\end{minipage}
\caption{(Color online) Free energy as a function of the shear component of the distortion tensor along the $xz$-plane for: a) $S_0 = 0.1$, $T = 0.3$, and several values of $\lambda$; b) $S_0 = 0.4$, $\lambda = 1.08$, and several temperatures. The transition is of second and first order in a) and b) respectively.}%
\label{free-energy}%
\end{figure*}

The free energy behavior near the second and first-order transitions in the phase diagram of Fig. \ref{phasediagram1} is illustrated in Fig \ref{free-energy}. In Fig. \ref{free-energy}a), we plot $f (\kappa) - f(0)$ as a function of $\kappa$, for $S_0 = 0.1$, $T = 0.3$, and several values of $\lambda$. Notice that the free energy has minima at $\kappa = 0$ and $\kappa \neq 0$, for $\lambda < \lambda_1$ and $\lambda > \lambda_1$ respectively. The transition from the uniaxial nematic phase along the $z$-direction to the soft phase is of second order, with the solution $\kappa \neq 0$ emerging continuously from zero at $\lambda = \lambda_1$. Under further stretching, the minimum free energy solution for $\kappa$ reaches a maximum, and then decreases continuously to zero at $\lambda = \lambda_2$, signaling another transition to a uniaxial nematic phase with orientational order along the $x$-direction. This type of behavior is reproduced for all $S_0$ at sufficiently low temperatures. In Fig. \ref{free-energy}b), we plot $f (\kappa) - f(0)$ as a function of $\kappa$, for $S_0 = 0.4$, $\lambda = 1.08$, and several temperatures. For a range of temperatures the free energy presents two local minima for the same $\lambda$, one at $\kappa = 0$ and the other at $\kappa \neq 0$. For $T > T_s$, the minimum at $\kappa \neq 0$ has a higher free energy, and the solution is metastable. This behavior is reversed for $T < T_s$, where the solution $\kappa = 0$ becomes metastable, so that the system is in the soft phase. The two solutions have the same free energy at $T = T_s$, characterizing a first-order phase transition.

\begin{figure}[!ht]
\vspace{0.5cm}
 \begin{center}
 \includegraphics[width=\linewidth]{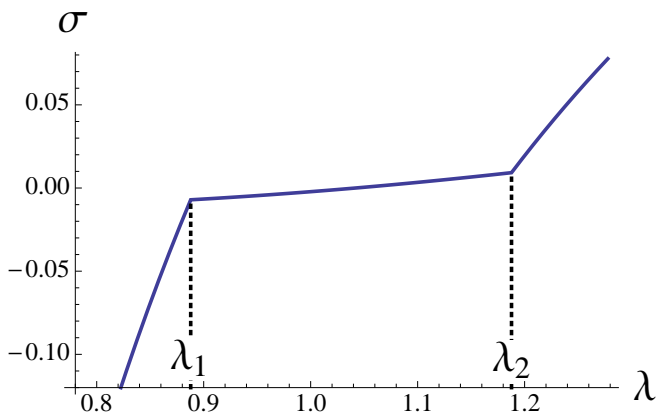}
 \end{center}
 \caption{(Color online) Engineering stress as a function of the distortion factor (along the $x$-direction), showing soft response for $\lambda_1 < \lambda < \lambda_2$.}
 \label{stressxdistortion}%
\end{figure}

\begin{figure}[!ht]
\vspace{0.5cm}
 \begin{center}
 \includegraphics[width=\linewidth]{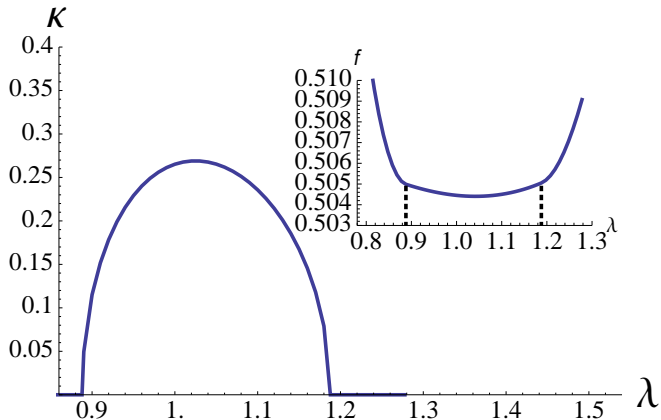}
 \end{center}
 \caption{(Color online) Shear component of the distortion tensor in the $xz$-plane, as a function of distortion  ($xx$ component). The free energy is shown in the inset.}
 \label{kappaandenergy}%
\end{figure}

At constant low-enough temperature, the engineering stress response is different in each of the three regions: $\lambda < \lambda_1$; $\lambda_1 < \lambda < \lambda_2$; and $\lambda > \lambda_2$ (Fig. \ref{stressxdistortion}). For $\lambda < \lambda_1$, the solution with $\kappa = 0$ minimizes the free energy, so that $Q_{xz} = 0$, and the state is uniaxial nematic with orientational order along the $z$-axis. For $\lambda_1 < \lambda < \lambda_2$, the solution with $\kappa \neq 0$ minimizes the free energy, the response is soft, with a very short slope in the $\lambda \times \sigma$ curve. As it happens, the nematic solution is also uniaxial over the whole soft region, since $Q$ always has two degenerate eigenvalues. Incidentally, the strain tensor is generally biaxial. The distortions $\lambda_1$ and $\lambda_2$ signal two second-order phase transitions. Within the soft phase, the shear component $\kappa$ increases continuously from zero at $\lambda_1$, reaches a maximum, and then decreases until it reaches zero at $\lambda_2$ (Fig. \ref{kappaandenergy}), and the free energy presents a shallow bottom (inset in Fig. \ref{kappaandenergy}). For $\lambda > \lambda_2$, the solution with $\kappa = 0$ minimizes the free energy again, $Q_{xz} = 0$, and the state is uniaxial nematic with orientational order along the $x$-axis. Notice that we explore the more comprehensive deformation range that includes sample compression, for $\lambda < \lambda_{\text{eq}}$, where $\lambda_{\text{eq}}$ is the value of $\lambda$ for which $\sigma = 0$. Below $\lambda_{\text{eq}}$ the nominal stress is naturally negative. If we restrict the plot to the region of positive nominal stress only, our results suggest a ``soft'', instead of ``semisoft'' response, according to the standard terminology. The same qualitative behavior is observed for all $S_0$ at sufficiently low temperatures.

\begin{figure}[!ht]
\vspace{0.5cm}
 \begin{center}
 \includegraphics[width=\linewidth]{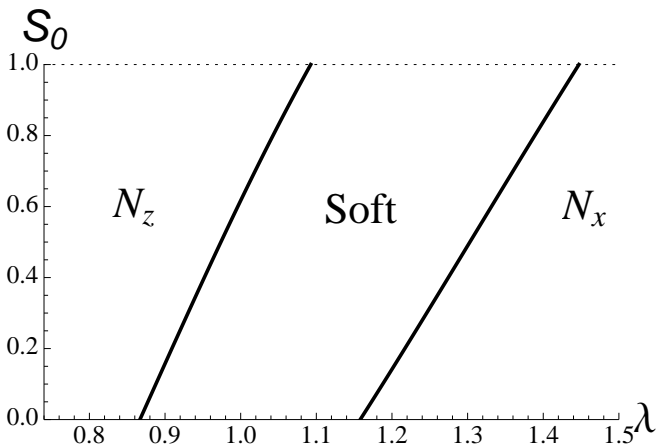}
 \end{center}
 \caption{Phase diagram in terms of the distortion factor and the initial scalar nematic order, for $T=0.3$.}
 \label{s0xlambda}%
\end{figure}

Fig. \ref{s0xlambda} displays a phase diagram in terms of the initial scalar nematic order and the distortion factor, for $T = 0.3$. The lines represent second-order phase transitions. Notice that $S_0$ yields only to a shift of the soft region. This topology is always found for sufficiently low temperatures, but might change in the neighborhood of one of the tricritical points.

\begin{figure}[!ht]
\vspace{0.5cm}
 \begin{center}
 \includegraphics[width=\linewidth]{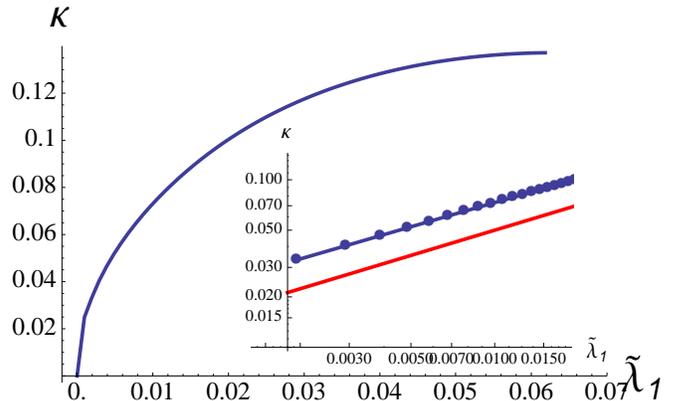}
 \end{center}
 \caption{(Color online) Shear component of the distortion tensor as a function of scaled distortion factor along the coexistence line. In the inset, the log-log plot suggests that $\kappa \sim \tilde{\lambda}_1^{1/2}$ close to the tricritical point.}
 \label{label7}%
\end{figure}

In Fig. \ref{label7} we plot $\kappa$ as a function of $\tilde{\lambda}_1$ along the coexistence line of Fig. \ref{phasediagram1}, where $S_0=0.4$, $\tilde{\lambda}_1 = |(\lambda- \lambda_1^\prime)/\lambda_1^\prime|$, and $\lambda_1^\prime$ is the tricritical point on the left. The inset shows a log-log plot of the same solution (blue dots), along with the line $y = \tilde{\lambda}^{1/2}$ (red full line online), suggesting that close to the tricritical point the shear component behaves approximately as $\kappa \sim \tilde{\lambda}_1^{1/2}$. This approximate exponent slightly changes depending on the tricritical point considered (on the left or right), and for different $S_0$ as well. It is interesting to speculate whether this behavior is expected to be found in finite-dimensional simulations and/or experiments. On the one hand, for the Ising model, with a few assumptions for the correlation function calculation, Ginzburg's criterium establishes that tricritical behavior has upper-critical dimension $d=3$ \cite{cardy96}. Roughly speaking, that means that close enough to tricritical points, predictions from mean-field theory are quantitatively reliable for realistic three-dimensional systems. On the other hand, the upper-critical dimension for the quenched random-field Ising model is six. Whereas a proper analysis of the Ginzburg criterium is still missing for our model, it would be interesting if the present work could stimulate further investigations on the tricritical behavior by numerical simulations and experiments. As far as the author know, the present work is the first paper reporting on the existence of tricritical points for transitions to the soft phase of nematic elastomers.

\section{Conclusions}

Along the lines of a previous work \cite{liarte11}, we introduce a lattice statistical model for the description of some peculiar features of soft nematic elastomers. This model includes a full continuum of orientations of the local directors, and suitable quantities to incorporate the initial nematic order. At the mean-field level, for a fully connected system, we use standard methods of statistical mechanics and numerical analysis to draw a phase diagram in terms of temperature and the distortion factor, which displays first- and second-order phase transitions, and the emergence of tricritical points. The soft phase is characterized by a non-vanishing shear component of the distortion tensor. Also, although our model yields to three distinct eigenvalues for the strain tensor, the nematic phase is still uniaxial in the soft region.

We hope that our analysis stimulates further investigation of the tricritical behavior of soft nematic elastomers, especially by means of numerical simulations of finite-dimensional systems and experiments. At last, we believe that a microscopic statistical model for elastomers should make room for inhomogeneities of the strain tensor, which have not been incorporated in our global spatially-uniform distortion tensor. These inhomogeneities are related to domain structures of real systems \cite{simone00, conti02a, conti02b}, which is a topic of capital interest for a future work on soft nematic elastomers.

\begin{acknowledgments}
I would like to acknowledge the financial support provided by the Brazilian agency Fapesp, and thank the Isaac Newton Institute for Mathematical Sciences, for their hospitality during the final stages of this work. Useful comments and insights from Silvio Salinas and James Sethna are also acknowledged.
\end{acknowledgments}



\begin{thebibliography}{37}%
\makeatletter
\providecommand \@ifxundefined [1]{%
 \@ifx{#1\undefined}
}%
\providecommand \@ifnum [1]{%
 \ifnum #1\expandafter \@firstoftwo
 \else \expandafter \@secondoftwo
 \fi
}%
\providecommand \@ifx [1]{%
 \ifx #1\expandafter \@firstoftwo
 \else \expandafter \@secondoftwo
 \fi
}%
\providecommand \natexlab [1]{#1}%
\providecommand \enquote  [1]{``#1''}%
\providecommand \bibnamefont  [1]{#1}%
\providecommand \bibfnamefont [1]{#1}%
\providecommand \citenamefont [1]{#1}%
\providecommand \href@noop [0]{\@secondoftwo}%
\providecommand \href [0]{\begingroup \@sanitize@url \@href}%
\providecommand \@href[1]{\@@startlink{#1}\@@href}%
\providecommand \@@href[1]{\endgroup#1\@@endlink}%
\providecommand \@sanitize@url [0]{\catcode `\\12\catcode `\$12\catcode
  `\&12\catcode `\#12\catcode `\^12\catcode `\_12\catcode `\%12\relax}%
\providecommand \@@startlink[1]{}%
\providecommand \@@endlink[0]{}%
\providecommand \url  [0]{\begingroup\@sanitize@url \@url }%
\providecommand \@url [1]{\endgroup\@href {#1}{\urlprefix }}%
\providecommand \urlprefix  [0]{URL }%
\providecommand \Eprint [0]{\href }%
\providecommand \doibase [0]{http://dx.doi.org/}%
\providecommand \selectlanguage [0]{\@gobble}%
\providecommand \bibinfo  [0]{\@secondoftwo}%
\providecommand \bibfield  [0]{\@secondoftwo}%
\providecommand \translation [1]{[#1]}%
\providecommand \BibitemOpen [0]{}%
\providecommand \bibitemStop [0]{}%
\providecommand \bibitemNoStop [0]{.\EOS\space}%
\providecommand \EOS [0]{\spacefactor3000\relax}%
\providecommand \BibitemShut  [1]{\csname bibitem#1\endcsname}%
\let\auto@bib@innerbib\@empty
\bibitem [{\citenamefont {de~Gennes}(1969)}]{gennes69}%
  \BibitemOpen
  \bibfield  {author} {\bibinfo {author} {\bibfnamefont {P.~G.}\ \bibnamefont
  {de~Gennes}},\ }\href@noop {} {\bibfield  {journal} {\bibinfo  {journal}
  {Phys. Lett.}\ }\textbf {\bibinfo {volume} {28A}},\ \bibinfo {pages} {725}
  (\bibinfo {year} {1969})}\BibitemShut {NoStop}%
\bibitem [{\citenamefont {Stenull}\ and\ \citenamefont
  {Lubensky}(2004)}]{stenull04}%
  \BibitemOpen
  \bibfield  {author} {\bibinfo {author} {\bibfnamefont {O.}~\bibnamefont
  {Stenull}}\ and\ \bibinfo {author} {\bibfnamefont {T.~C.}\ \bibnamefont
  {Lubensky}},\ }\href@noop {} {\bibfield  {journal} {\bibinfo  {journal}
  {Phys. Rev. E}\ }\textbf {\bibinfo {volume} {69}},\ \bibinfo {pages} {021807}
  (\bibinfo {year} {2004})}\BibitemShut {NoStop}%
\bibitem [{\citenamefont {Ennis}\ \emph {et~al.}(2006)\citenamefont {Ennis},
  \citenamefont {Malacarne}, \citenamefont {Palffy-Muhoray},\ and\
  \citenamefont {Shelley}}]{ennis06}%
  \BibitemOpen
  \bibfield  {author} {\bibinfo {author} {\bibfnamefont {R.}~\bibnamefont
  {Ennis}}, \bibinfo {author} {\bibfnamefont {L.~C.}\ \bibnamefont
  {Malacarne}}, \bibinfo {author} {\bibfnamefont {P.}~\bibnamefont
  {Palffy-Muhoray}}, \ and\ \bibinfo {author} {\bibfnamefont {M.}~\bibnamefont
  {Shelley}},\ }\href@noop {} {\bibfield  {journal} {\bibinfo  {journal} {Phys.
  Rev. E}\ }\textbf {\bibinfo {volume} {74}},\ \bibinfo {pages} {061802}
  (\bibinfo {year} {2006})}\BibitemShut {NoStop}%
\bibitem [{\citenamefont {Ye}\ \emph {et~al.}(2007)\citenamefont {Ye},
  \citenamefont {Mukhopadhyay}, \citenamefont {Stenull},\ and\ \citenamefont
  {Lubensky}}]{ye07}%
  \BibitemOpen
  \bibfield  {author} {\bibinfo {author} {\bibfnamefont {F.}~\bibnamefont
  {Ye}}, \bibinfo {author} {\bibfnamefont {R.}~\bibnamefont {Mukhopadhyay}},
  \bibinfo {author} {\bibfnamefont {O.}~\bibnamefont {Stenull}}, \ and\
  \bibinfo {author} {\bibfnamefont {T.~C.}\ \bibnamefont {Lubensky}},\
  }\href@noop {} {\bibfield  {journal} {\bibinfo  {journal} {Phys. Rev. Lett.}\
  }\textbf {\bibinfo {volume} {98}},\ \bibinfo {pages} {147801} (\bibinfo
  {year} {2007})}\BibitemShut {NoStop}%
\bibitem [{\citenamefont {Zhu}\ \emph {et~al.}(2011)\citenamefont {Zhu},
  \citenamefont {Shelley},\ and\ \citenamefont {Palffy-Muhoray}}]{zhu11}%
  \BibitemOpen
  \bibfield  {author} {\bibinfo {author} {\bibfnamefont {W.}~\bibnamefont
  {Zhu}}, \bibinfo {author} {\bibfnamefont {M.}~\bibnamefont {Shelley}}, \ and\
  \bibinfo {author} {\bibfnamefont {P.}~\bibnamefont {Palffy-Muhoray}},\
  }\href@noop {} {\bibfield  {journal} {\bibinfo  {journal} {Phys. Rev. E}\
  }\textbf {\bibinfo {volume} {83}},\ \bibinfo {pages} {051703} (\bibinfo
  {year} {2011})}\BibitemShut {NoStop}%
\bibitem [{\citenamefont {Pasini}\ \emph {et~al.}(2005)\citenamefont {Pasini},
  \citenamefont {Ska\v{c}ej},\ and\ \citenamefont {Zannoni}}]{pasini05}%
  \BibitemOpen
  \bibfield  {author} {\bibinfo {author} {\bibfnamefont {P.}~\bibnamefont
  {Pasini}}, \bibinfo {author} {\bibfnamefont {G.}~\bibnamefont {Ska\v{c}ej}},
  \ and\ \bibinfo {author} {\bibfnamefont {C.}~\bibnamefont {Zannoni}},\
  }\href@noop {} {\bibfield  {journal} {\bibinfo  {journal} {Chem. Phys.
  Lett.}\ }\textbf {\bibinfo {volume} {413}},\ \bibinfo {pages} {463} (\bibinfo
  {year} {2005})}\BibitemShut {NoStop}%
\bibitem [{\citenamefont {Ska\u{c}ej}\ and\ \citenamefont
  {Zannoni}(2011)}]{skacej11}%
  \BibitemOpen
  \bibfield  {author} {\bibinfo {author} {\bibfnamefont {G.}~\bibnamefont
  {Ska\u{c}ej}}\ and\ \bibinfo {author} {\bibfnamefont {C.}~\bibnamefont
  {Zannoni}},\ }\href@noop {} {\bibfield  {journal} {\bibinfo  {journal} {Soft
  Matter}\ }\textbf {\bibinfo {volume} {7}},\ \bibinfo {pages} {9983} (\bibinfo
  {year} {2011})}\BibitemShut {NoStop}%
\bibitem [{\citenamefont {Whitmer}\ \emph {et~al.}(2013)\citenamefont
  {Whitmer}, \citenamefont {Roberts}, \citenamefont {Shekhar}, \citenamefont
  {Abbott},\ and\ \citenamefont {de~Pablo}}]{whitmer13}%
  \BibitemOpen
  \bibfield  {author} {\bibinfo {author} {\bibfnamefont {J.~K.}\ \bibnamefont
  {Whitmer}}, \bibinfo {author} {\bibfnamefont {T.~F.}\ \bibnamefont
  {Roberts}}, \bibinfo {author} {\bibfnamefont {R.}~\bibnamefont {Shekhar}},
  \bibinfo {author} {\bibfnamefont {N.~L.}\ \bibnamefont {Abbott}}, \ and\
  \bibinfo {author} {\bibfnamefont {J.~J.}\ \bibnamefont {de~Pablo}},\
  }\href@noop {} {\bibfield  {journal} {\bibinfo  {journal} {Phys. Rev. E}\
  }\textbf {\bibinfo {volume} {87}},\ \bibinfo {pages} {020502(R)} (\bibinfo
  {year} {2013})}\BibitemShut {NoStop}%
\bibitem [{\citenamefont {Selinger}\ and\ \citenamefont
  {Ratna}(2004)}]{selinger04}%
  \BibitemOpen
  \bibfield  {author} {\bibinfo {author} {\bibfnamefont {J.~V.}\ \bibnamefont
  {Selinger}}\ and\ \bibinfo {author} {\bibfnamefont {B.~R.}\ \bibnamefont
  {Ratna}},\ }\href@noop {} {\bibfield  {journal} {\bibinfo  {journal} {Phys.
  Rev. E}\ }\textbf {\bibinfo {volume} {70}},\ \bibinfo {pages} {041707}
  (\bibinfo {year} {2004})}\BibitemShut {NoStop}%
\bibitem [{\citenamefont {Xing}\ \emph {et~al.}(2008)\citenamefont {Xing},
  \citenamefont {Pfahl}, \citenamefont {Mukhopadhyay}, \citenamefont
  {Goldbart},\ and\ \citenamefont {Zippelius}}]{xing08}%
  \BibitemOpen
  \bibfield  {author} {\bibinfo {author} {\bibfnamefont {X.}~\bibnamefont
  {Xing}}, \bibinfo {author} {\bibfnamefont {S.}~\bibnamefont {Pfahl}},
  \bibinfo {author} {\bibfnamefont {S.}~\bibnamefont {Mukhopadhyay}}, \bibinfo
  {author} {\bibfnamefont {P.~M.}\ \bibnamefont {Goldbart}}, \ and\ \bibinfo
  {author} {\bibfnamefont {A.}~\bibnamefont {Zippelius}},\ }\href@noop {}
  {\bibfield  {journal} {\bibinfo  {journal} {Phys. Rev. E}\ }\textbf {\bibinfo
  {volume} {77}},\ \bibinfo {pages} {051802} (\bibinfo {year}
  {2008})}\BibitemShut {NoStop}%
\bibitem [{\citenamefont {Liarte}\ \emph {et~al.}(2011)\citenamefont {Liarte},
  \citenamefont {Salinas},\ and\ \citenamefont {Yokoi}}]{liarte11}%
  \BibitemOpen
  \bibfield  {author} {\bibinfo {author} {\bibfnamefont {D.~B.}\ \bibnamefont
  {Liarte}}, \bibinfo {author} {\bibfnamefont {S.~R.}\ \bibnamefont {Salinas}},
  \ and\ \bibinfo {author} {\bibfnamefont {C.~S.~O.}\ \bibnamefont {Yokoi}},\
  }\href@noop {} {\bibfield  {journal} {\bibinfo  {journal} {Phys. Rev. E}\
  }\textbf {\bibinfo {volume} {84}},\ \bibinfo {pages} {011124} (\bibinfo
  {year} {2011})}\BibitemShut {NoStop}%
\bibitem [{\citenamefont {Bladon}\ \emph {et~al.}(1994)\citenamefont {Bladon},
  \citenamefont {Terentjev},\ and\ \citenamefont {Warner}}]{bladon94}%
  \BibitemOpen
  \bibfield  {author} {\bibinfo {author} {\bibfnamefont {P.}~\bibnamefont
  {Bladon}}, \bibinfo {author} {\bibfnamefont {E.~M.}\ \bibnamefont
  {Terentjev}}, \ and\ \bibinfo {author} {\bibfnamefont {M.}~\bibnamefont
  {Warner}},\ }\href@noop {} {\bibfield  {journal} {\bibinfo  {journal} {J.
  Phys. II (France)}\ }\textbf {\bibinfo {volume} {4}},\ \bibinfo {pages} {75}
  (\bibinfo {year} {1994})}\BibitemShut {NoStop}%
\bibitem [{\citenamefont {Warner}\ and\ \citenamefont
  {Terentjev}(1996)}]{warner96}%
  \BibitemOpen
  \bibfield  {author} {\bibinfo {author} {\bibfnamefont {M.}~\bibnamefont
  {Warner}}\ and\ \bibinfo {author} {\bibfnamefont {E.~M.}\ \bibnamefont
  {Terentjev}},\ }\href@noop {} {\bibfield  {journal} {\bibinfo  {journal}
  {Prog. Polym. Sci.}\ }\textbf {\bibinfo {volume} {21}},\ \bibinfo {pages}
  {853} (\bibinfo {year} {1996})}\BibitemShut {NoStop}%
\bibitem [{\citenamefont {Warner}\ and\ \citenamefont
  {Terentjev}(2003)}]{warner03}%
  \BibitemOpen
  \bibfield  {author} {\bibinfo {author} {\bibfnamefont {M.}~\bibnamefont
  {Warner}}\ and\ \bibinfo {author} {\bibfnamefont {E.~M.}\ \bibnamefont
  {Terentjev}},\ }\href@noop {} {\emph {\bibinfo {title} {Liquid Crystal
  Elastomers}}}\ (\bibinfo  {publisher} {Oxford University Press, Oxford},\
  \bibinfo {year} {2003})\BibitemShut {NoStop}%
\bibitem [{\citenamefont {Lebwohl}\ and\ \citenamefont
  {Lasher}(1972)}]{lebwohl72}%
  \BibitemOpen
  \bibfield  {author} {\bibinfo {author} {\bibfnamefont {P.~A.}\ \bibnamefont
  {Lebwohl}}\ and\ \bibinfo {author} {\bibfnamefont {G.}~\bibnamefont
  {Lasher}},\ }\href@noop {} {\bibfield  {journal} {\bibinfo  {journal} {Phys.
  Rev. A}\ }\textbf {\bibinfo {volume} {6}},\ \bibinfo {pages} {426} (\bibinfo
  {year} {1972})}\BibitemShut {NoStop}%
\bibitem [{Note1()}]{Note1}%
  \BibitemOpen
  \bibinfo {note} {The usual distinction between ``soft'' and ``semi-soft''
  responses associates the later with the existence of Hookean behavior at the
  beginning of the stress-strain curve.}\BibitemShut {Stop}%
\bibitem [{\citenamefont {Warner}\ and\ \citenamefont
  {Kutter}(2002)}]{warner02}%
  \BibitemOpen
  \bibfield  {author} {\bibinfo {author} {\bibfnamefont {M.}~\bibnamefont
  {Warner}}\ and\ \bibinfo {author} {\bibfnamefont {S.}~\bibnamefont
  {Kutter}},\ }\href@noop {} {\bibfield  {journal} {\bibinfo  {journal} {Phys.
  Rev. E}\ }\textbf {\bibinfo {volume} {65}},\ \bibinfo {pages} {051707}
  (\bibinfo {year} {2002})}\BibitemShut {NoStop}%
\bibitem [{\citenamefont {Ye}\ and\ \citenamefont {Lubensky}(2009)}]{ye09}%
  \BibitemOpen
  \bibfield  {author} {\bibinfo {author} {\bibfnamefont {F.}~\bibnamefont
  {Ye}}\ and\ \bibinfo {author} {\bibfnamefont {T.~C.}\ \bibnamefont
  {Lubensky}},\ }\href@noop {} {\bibfield  {journal} {\bibinfo  {journal} {J.
  Phys. Chem. B}\ }\textbf {\bibinfo {volume} {113}},\ \bibinfo {pages} {3853}
  (\bibinfo {year} {2009})}\BibitemShut {NoStop}%
\bibitem [{Note2()}]{Note2}%
  \BibitemOpen
  \bibinfo {note} {There is no transition between Nz and Nx. They belong to the
  same standard uniaxial phase. The optical axis continuously rotates from the
  z- to the x-direction upon increasing $\lambda $. Also, there is no
  nematic-isotropic transition, since the imposed strain acts as an aligning
  mechanical field.}\BibitemShut {Stop}%
\bibitem [{\citenamefont {Maier}\ and\ \citenamefont {Saupe}(1958)}]{maier58}%
  \BibitemOpen
  \bibfield  {author} {\bibinfo {author} {\bibfnamefont {W.}~\bibnamefont
  {Maier}}\ and\ \bibinfo {author} {\bibfnamefont {A.}~\bibnamefont {Saupe}},\
  }\href@noop {} {\bibfield  {journal} {\bibinfo  {journal} {Z. Naturforsch. A:
  Phys. Sci.}\ }\textbf {\bibinfo {volume} {13}},\ \bibinfo {pages} {564}
  (\bibinfo {year} {1958})}\BibitemShut {NoStop}%
\bibitem [{Note3()}]{Note3}%
  \BibitemOpen
  \bibinfo {note} {With appropriate choices for $A$ and the zero-point energy,
  this model can be mapped into the Lebwohl-Lasher model \cite {lebwohl72}.
  Later on we will take the mean-field approximation by considering
  infinite-range interactions between mesogens \cite
  {kac68,stanley71,salinas93}, so that the initial lattice topology becomes
  irrelevant.}\BibitemShut {Stop}%
\bibitem [{\citenamefont {Landau}\ and\ \citenamefont
  {Lifshitz}(1986)}]{landau86}%
  \BibitemOpen
  \bibfield  {author} {\bibinfo {author} {\bibfnamefont {L.~D.}\ \bibnamefont
  {Landau}}\ and\ \bibinfo {author} {\bibfnamefont {E.~M.}\ \bibnamefont
  {Lifshitz}},\ }\href@noop {} {\emph {\bibinfo {title} {Theory of
  Elasticity}}},\ \bibinfo {edition} {3rd}\ ed.\ (\bibinfo  {publisher}
  {Elsevier, New York},\ \bibinfo {year} {1986})\BibitemShut {NoStop}%
\bibitem [{\citenamefont {Chaikin}\ and\ \citenamefont
  {Lubensky}(1995)}]{chaikin95}%
  \BibitemOpen
  \bibfield  {author} {\bibinfo {author} {\bibfnamefont {P.~M.}\ \bibnamefont
  {Chaikin}}\ and\ \bibinfo {author} {\bibfnamefont {T.~C.}\ \bibnamefont
  {Lubensky}},\ }\href@noop {} {\emph {\bibinfo {title} {Principles of
  condensed matter physics}}}\ (\bibinfo  {publisher} {Cambridge University
  Press, Cambridge},\ \bibinfo {year} {1995})\BibitemShut {NoStop}%
\bibitem [{Note4()}]{Note4}%
  \BibitemOpen
  \bibinfo {note} {Notice that $b=0$ for isotropic rubber. Also, $a=(2l_{\perp
  }^{-1} + l_{\parallel }^{-1})/3$, and $b=2(l_{\parallel }^{-1}-l_{\perp
  }^{-1})/3$, in terms of the parallel ($l_\parallel $) and perpendicular
  ($l_{\perp }$) effective displacements of the polymer chain \cite
  {liarte11,warner03}. Information about the type of nematic polymer (main
  chain and side chain) is condensed in these parameters. In general, the ratio
  $l_{\parallel } / l_{\perp }$ is higher for main chain nematic polymers,
  leading to more extreme mechanical effects (see section 3.2.1 of \cite
  {warner03})}\BibitemShut {NoStop}%
\bibitem [{\citenamefont {Kac}(1968)}]{kac68}%
  \BibitemOpen
  \bibfield  {author} {\bibinfo {author} {\bibfnamefont {M.}~\bibnamefont
  {Kac}},\ }in\ \href@noop {} {\emph {\bibinfo {booktitle} {Statistical
  physics, phase transitions, and superfluidity}}},\ \bibinfo {editor} {edited
  by\ \bibinfo {editor} {\bibfnamefont {M.}~\bibnamefont {Chr\'etien}},
  \bibinfo {editor} {\bibfnamefont {E.~P.}\ \bibnamefont {Gross}}, \ and\
  \bibinfo {editor} {\bibfnamefont {S.}~\bibnamefont {Deser}}}\ (\bibinfo
  {publisher} {Gordon and Breach, New York},\ \bibinfo {year} {1968})\ p.\
  \bibinfo {pages} {241}\BibitemShut {NoStop}%
\bibitem [{\citenamefont {Stanley}(1971)}]{stanley71}%
  \BibitemOpen
  \bibfield  {author} {\bibinfo {author} {\bibfnamefont {H.~G.}\ \bibnamefont
  {Stanley}},\ }\href@noop {} {\emph {\bibinfo {title} {Phase Transition and
  Critical Phenomena}}}\ (\bibinfo  {publisher} {Oxford University Press,
  Oxford},\ \bibinfo {year} {1971})\BibitemShut {NoStop}%
\bibitem [{\citenamefont {Salinas}\ and\ \citenamefont
  {Wreszinski}(1993)}]{salinas93}%
  \BibitemOpen
  \bibfield  {author} {\bibinfo {author} {\bibfnamefont {S.~R.}\ \bibnamefont
  {Salinas}}\ and\ \bibinfo {author} {\bibfnamefont {W.~F.}\ \bibnamefont
  {Wreszinski}},\ }\href@noop {} {\emph {\bibinfo {title} {Disorder and
  Competition in Soluble Lattice Models}}}\ (\bibinfo  {publisher} {World
  Scientific, Singapore},\ \bibinfo {year} {1993})\BibitemShut {NoStop}%
\bibitem [{\citenamefont {Henriques}\ and\ \citenamefont
  {Henriques}(1997)}]{henriques97}%
  \BibitemOpen
  \bibfield  {author} {\bibinfo {author} {\bibfnamefont {E.~F.}\ \bibnamefont
  {Henriques}}\ and\ \bibinfo {author} {\bibfnamefont {V.~B.}\ \bibnamefont
  {Henriques}},\ }\href@noop {} {\bibfield  {journal} {\bibinfo  {journal} {J.
  Chem. Phys.}\ }\textbf {\bibinfo {volume} {107}},\ \bibinfo {pages} {8036}
  (\bibinfo {year} {1997})}\BibitemShut {NoStop}%
\bibitem [{\citenamefont {do~Carmo}\ \emph {et~al.}(2010)\citenamefont
  {do~Carmo}, \citenamefont {Liarte},\ and\ \citenamefont {Salinas}}]{carmo10}%
  \BibitemOpen
  \bibfield  {author} {\bibinfo {author} {\bibfnamefont {E.}~\bibnamefont
  {do~Carmo}}, \bibinfo {author} {\bibfnamefont {D.~B.}\ \bibnamefont
  {Liarte}}, \ and\ \bibinfo {author} {\bibfnamefont {S.~R.}\ \bibnamefont
  {Salinas}},\ }\href@noop {} {\bibfield  {journal} {\bibinfo  {journal} {Phys.
  Rev. E}\ }\textbf {\bibinfo {volume} {81}},\ \bibinfo {pages} {062701}
  (\bibinfo {year} {2010})}\BibitemShut {NoStop}%
\bibitem [{\citenamefont {Liarte}\ and\ \citenamefont
  {Salinas}(2012)}]{liarte12}%
  \BibitemOpen
  \bibfield  {author} {\bibinfo {author} {\bibfnamefont {D.~B.}\ \bibnamefont
  {Liarte}}\ and\ \bibinfo {author} {\bibfnamefont {S.~R.}\ \bibnamefont
  {Salinas}},\ }\href@noop {} {\bibfield  {journal} {\bibinfo  {journal} {Braz.
  J. Phys.}\ }\textbf {\bibinfo {volume} {42}},\ \bibinfo {pages} {261}
  (\bibinfo {year} {2012})}\BibitemShut {NoStop}%
\bibitem [{\citenamefont {Zwanzig}(1963)}]{zwanzig63}%
  \BibitemOpen
  \bibfield  {author} {\bibinfo {author} {\bibfnamefont {R.}~\bibnamefont
  {Zwanzig}},\ }\href@noop {} {\bibfield  {journal} {\bibinfo  {journal} {J.
  Chem. Phys.}\ }\textbf {\bibinfo {volume} {39}},\ \bibinfo {pages} {1714}
  (\bibinfo {year} {1963})}\BibitemShut {NoStop}%
\bibitem [{Note5()}]{Note5}%
  \BibitemOpen
  \bibinfo {note} {Whenever possible we use Einstein summation rule for
  repeated indices in order to simplify notation.}\BibitemShut {Stop}%
\bibitem [{\citenamefont {Marsden}\ and\ \citenamefont
  {Hughes}(1968)}]{marsden68}%
  \BibitemOpen
  \bibfield  {author} {\bibinfo {author} {\bibfnamefont {J.~E.}\ \bibnamefont
  {Marsden}}\ and\ \bibinfo {author} {\bibfnamefont {T.~J.}\ \bibnamefont
  {Hughes}},\ }\href@noop {} {\emph {\bibinfo {title} {Mathematical Foundations
  of Elasticity}}}\ (\bibinfo  {publisher} {Printice-Hall, Inc.: Englewood
  Cliffs, NJ},\ \bibinfo {year} {1968})\BibitemShut {NoStop}%
\bibitem [{\citenamefont {Cardy}(1996)}]{cardy96}%
  \BibitemOpen
  \bibfield  {author} {\bibinfo {author} {\bibfnamefont {J.}~\bibnamefont
  {Cardy}},\ }\href@noop {} {\emph {\bibinfo {title} {Scaling and
  Renormalization in Statistical Physics}}}\ (\bibinfo  {publisher} {Cambridge
  University Press, Cambridge},\ \bibinfo {year} {1996})\BibitemShut {NoStop}%
\bibitem [{\citenamefont {de~Simone}\ and\ \citenamefont
  {Dolzmann}(2000)}]{simone00}%
  \BibitemOpen
  \bibfield  {author} {\bibinfo {author} {\bibfnamefont {A.}~\bibnamefont
  {de~Simone}}\ and\ \bibinfo {author} {\bibfnamefont {G.}~\bibnamefont
  {Dolzmann}},\ }\href@noop {} {\bibfield  {journal} {\bibinfo  {journal}
  {Physica D}\ }\textbf {\bibinfo {volume} {136}},\ \bibinfo {pages} {175}
  (\bibinfo {year} {2000})}\BibitemShut {NoStop}%
\bibitem [{\citenamefont {Conti}\ \emph
  {et~al.}(2002{\natexlab{a}})\citenamefont {Conti}, \citenamefont
  {de~Simone},\ and\ \citenamefont {Dolzmann}}]{conti02a}%
  \BibitemOpen
  \bibfield  {author} {\bibinfo {author} {\bibfnamefont {S.}~\bibnamefont
  {Conti}}, \bibinfo {author} {\bibfnamefont {A.}~\bibnamefont {de~Simone}}, \
  and\ \bibinfo {author} {\bibfnamefont {G.}~\bibnamefont {Dolzmann}},\
  }\href@noop {} {\bibfield  {journal} {\bibinfo  {journal} {J. Mech. Phys.
  Solids}\ }\textbf {\bibinfo {volume} {50}},\ \bibinfo {pages} {1431}
  (\bibinfo {year} {2002}{\natexlab{a}})}\BibitemShut {NoStop}%
\bibitem [{\citenamefont {Conti}\ \emph
  {et~al.}(2002{\natexlab{b}})\citenamefont {Conti}, \citenamefont
  {de~Simone},\ and\ \citenamefont {Dolzmann}}]{conti02b}%
  \BibitemOpen
  \bibfield  {author} {\bibinfo {author} {\bibfnamefont {S.}~\bibnamefont
  {Conti}}, \bibinfo {author} {\bibfnamefont {A.}~\bibnamefont {de~Simone}}, \
  and\ \bibinfo {author} {\bibfnamefont {G.}~\bibnamefont {Dolzmann}},\
  }\href@noop {} {\bibfield  {journal} {\bibinfo  {journal} {Phys. Rev. E}\
  }\textbf {\bibinfo {volume} {66}},\ \bibinfo {pages} {061710} (\bibinfo
  {year} {2002}{\natexlab{b}})}\BibitemShut {NoStop}%
\end{thebibliography}
%
\end{document}